\documentclass[%
reprint, twocolumn,
superscriptaddress,
showpacs,preprintnumbers,
nofootinbib,
amsmath,amssymb, aps, pra,
]{revtex4}

\usepackage{amsmath}
\usepackage{mathrsfs}
\usepackage{calrsfs}
\usepackage{multirow}
\DeclareMathAlphabet{\pazocal}{OMS}{zplm}{m}{n}

\usepackage{graphicx}
\usepackage{dcolumn}
\usepackage{bm}
\usepackage{color} 
\usepackage{CJK}
\usepackage{dsfont}
\usepackage[extension=xxx]{hyperref}

\newcommand{\beq}{\begin{equation}}
\newcommand{\eeq}{\end{equation}}
\newcommand{\beqa}{\begin{eqnarray}}
\newcommand{\eeqa}{\end{eqnarray}}

\def\beq{\begin{equation}}

\begin{document}
\title{Invariant-based inverse engineering of crane control parameters
}
\author{S. Gonz\'alez-Resines}
\affiliation{Departamento de Qu\'{\i}mica F\'{\i}sica, UPV/EHU, Apdo
644, 48080 Bilbao, Spain}
\author{D. Gu\'ery-Odelin}
\affiliation{Laboratoire de Collisions Agr\'egats R\'eactivit\'e, CNRS UMR 5589, IRSAMC, Universit\'e de Toulouse (UPS), 118 Route de Narbonne, 31062 Toulouse CEDEX 4, France}
\author{A. Tobalina}
\affiliation{Departamento de Qu\'{\i}mica F\'{\i}sica, UPV/EHU, Apdo
644, 48080 Bilbao, Spain}
\author{I. Lizuain}
\affiliation{Department of Applied Mathematics, University of the Basque Country UPV/EHU, Plaza Europa 1, 20018 Donostia-San Sebastian, Spain}
\author{E. Torrontegui}
\affiliation{Instituto de F\'{\i}sica Fundamental IFF-CSIC, Calle Serrano 113b, 28006 Madrid, Spain}
\author{J. G. Muga}
\affiliation{Departamento de Qu\'{\i}mica F\'{\i}sica, UPV/EHU, Apdo
644, 48080 Bilbao, Spain}

\date{\today}
\begin{abstract}
By applying invariant-based inverse engineering in the small-oscillations regime, 
we design the time dependence of the control parameters of an overhead crane (trolley displacement and rope length), to transport a load between two
positions at different heights with  minimal final energy excitation for a microcanonical ensemble of initial conditions.  The analogies between ion transport in multisegmented traps 
or neutral atom transport in moving optical lattices and load manipulation by cranes opens a route for a 
useful transfer of techniques among very different fields.    
\end{abstract}
\maketitle
\section{Introduction}
The similarities between the mathematical descriptions of different systems have been repeatedly applied in the history 
of Physics and continue to play a heuristic and  fruitful role.
The entire field of quantum simulations \cite{simu0,simu1,simu2,simu3}, or the design of optical waveguide devices based on analogies with 
quantum mechanical discrete systems \cite{optics1,optics2,optics3,optics4,optics5} are current  examples. The usefulness of analogies was already  
pointed out  by Maxwell \cite{Maxwell}: ``In many cases the relations of the phenomena in two
different physical questions have a certain similarity which enables
us, when we have solved one of these questions, to make use of our
solution in answering the other''. He also warned about the
dangers of pushing the analogy beyond certain limits, and the need of a ``judicious use''  \cite{Maxwell}. 

Our first aim in this work is to point out a connection between two very different systems so as to,
following Maxwell's advise, 
judiciously take advantage of the similarities for mutual benefit, by transferring  
knowledge and control techniques. Specifically the systems in question are   
{\it i)} ultracold  ions or neutral atoms in effectively one dimensional (1D) traps with time-dependent controllable parameters, as used in recent advances towards scalable quantum information processing \cite{Wineland,Kielpinski},
and {\it ii)} payloads moved by industrial or construction cranes \cite{review}. 
In spite of the different orders of magnitude in masses and sizes involved, both systems are described in the harmonic limit of small oscillations by 
similar mass-independent equations, and require manipulation approaches to achieve the same  goals, namely,  to transfer a mass quickly, safely, and without final excitation, to a preselected location. Common is also the need to 
implement robust protocols with respect to on route perturbations  or initial-state dispersion. 
Quantumness may play a role beyond the harmonic domain,  
and precludes a naive direct translation from the macroscopic to the microscopic worlds of control techniques based on feedback from measurements performed on 
route. However, in  the harmonic regime,   
the equations used for inverse engineering the control parameters in open-loop 
approaches (without feedback control) of a crane can be identical to those used to determine the motion of traps that drive microscopic particles. These approaches may as well have common elements for  quantum and classical systems beyond the harmonic domain. 
This paper primarily proposes a transfer of some results and inverse engineering techniques used to transport or launch trapped ions
and neutral atoms -shortcuts to adiabaticity \cite{reviewSTA2013}- 
to design control operations for cranes. By establishing the link among the two fields, we also open the way for 
a reverse transfer, from the considerable body of work and results developed by engineers to control crane
operation  onto the microscopic realm. Some  examples of transfer are discussed in the final section. In the rest of the introduction
we provide basic facts on cranes,  shortcuts to adiabaticity, and their application to cranes based on invariants of motion.

{\it Cranes.} Cranes are mechanical machines to lift and transport loads by means of a hoisting  rope supported by a structure that moves the suspension point.    
They range from gigantic proportions in ports or special construction sites to minicranes and even microscopic, molecular sized, devices \cite{moleccrane}.
Exact forms and types (overhead, rotary, or boom cranes \cite{review}) also vary widely to adapt to different applications and they can be operated manually or automatically.   
In either case standard objectives of the crane operation are to transport the load in a short time 
depositing it safely and at rest or at least unexcited at the final destination. Typically the residual (final) pendulations are to be minimized while larger pendulations are allowed on route as long as the safety of the operation is not at risk. 
The dynamical models are frequently linearized to apply linear control methods, 
as normal operation conditions involve small sway angles \cite{review,xia,so1,so2,so3,so4}.   
Open-loop (without feedback control) and closed loop (with feedback provided by sensors) strategies have been proposed. The potential superiority of closed loop approaches is  overtaken by their difficult implementation which requires accurate measurements during the operation, so that open-loop approaches are  dominant in actual crane controllers \cite{review}.  
Nevertheless, known problems of the open-loop approaches are their sensitivity with respect to different perturbations 
such as wind, damping, changes in initial conditions, or imperfect implementation of the control functions \cite{review,init}.   
Moreover, optimal solutions often imply bang-bang (stepwise constant) acceleration profiles of control parameters, which are hard  to implement and generate stress on the structure \cite{review}.

{\it Shortcuts to adiabaticity.} Slow transformations of external parameters that control quantum systems allow in principle 
to avoid excitations and in particular to reach ground states  that could be difficult to
achieve otherwise. These slow processes  are in principle robust with respect to smooth changes of parameter paths but not necessarily with respect to rapid or noisy perturbations. 
There is a growing interest in accelerating these transformations to limit the detrimental effect of decoherence or noise, or to  increase the number of operations that can be carried out in a given time interval \cite{reviewSTA2013}. 
Shortcuts to adiabaticity are methods that bypass slow adiabatic transformations via faster routes by designing 
appropriate time-dependent Hamiltonians. This includes methods based on 
exact solutions with scaling parameters \cite{DGO2009,PRL10,boltz14,PRA2014},  
and on dynamical Lewis-Riesenfeld invariants \cite{PRL10,NJP2012,PRL13}, 
approaches that add ``counterdiabatic'' terms to the Hamiltonian \cite{Rice,Berry,PRL2}, 
the ``fast-forward'' approach \cite{ffw1,ffw2,torron12,torron13,MNC14,Kazu1,Deffner16,Jar}, Lie-algebraic methods \cite{Lie1,Lie2}, 
or the fast quasi-adiabatic approach \cite{FAQUAD}. They all
have applications beyond the quantum 
domain, see e. g. applications in optics \cite{optics1,optics2,optics3,optics4,optics5}, and classical systems 
\cite{boltz14,cla1,cla2,cla3,Jar,Kazu}. 

{\it Invariants and cranes.}
Use of invariants of motion to inverse engineer the trolley trajectory and the rope length will allow us to design 
for the  open-loop strategy (for closed-loop control see \cite{init}) 
robust, smooth protocols with respect to initial conditions: specifically we shall generate a family of protocols that by construction produce at final time the adiabatic energy, namely, the energy that would be reached after an infinitely slow process.  The fact that the solution for the time dependence of the protocol parameters 
is highly degenerate allows for further optimization, with respect to other physical variables or robustness,  as demonstrated e.g. in \cite{NJP2012,noise} 
within the microscopic domain. We demonstrate here the possibility to use the degeneracy to devise protocols 
that remain robust  beyond the harmonic regime. 

We shall also establish a minimal work principle for the system,  which  generally states that on average, the adiabatic work  done on the system or extracted from it for very slow processes is optimal.
A brief history  of the principle and its various formulations and derivations for specific systems and conditions is provided in \cite{mwp1,mwp2}. 
For the present context we shall prove it  in Appendix \ref{aa}, considering a microcanonical ensemble
of initial conditions of the payload,  by generalizing the results in \cite{adiabinv}. The relevance of the result is that    
invariant-based inverse engineering of the crane control parameters provides for the microcanonical ensemble of initial conditions the minimal possible final average energy with faster-than-adiabatic protocols.

Section \ref{ibe}  explains how to implement invariant-based inverse engineering for crane control, 
and Sec. \ref{ne} provides some examples, including a comparison of sequential versus dual approaches. The article ends with a discussion containing an outlook for future work and a demonstration of the possibility to go beyond the harmonic regime. Finally, appendices are included on: 
the maximal work principle, the derivation of the Hamiltonian in the small oscillations regime, 
and the expression for the exact adiabatic energy. 
\section{Invariant-based engineering of crane control\label{ibe}}
Invariant-based engineering is the approach best adapted to the peculiarities of trapped ions and has been extensively applied by our group in that context \cite{Pal1,Pal2,Pal3,Pal4,Pal5,Pal6,Pal7,Lizuain}.
Here we propose  an inverse engineering method for crane operations based on the invariants of motion of the system. While the approach can be generalized and applied to different, complex  crane types, to be specific we set a simple overhead crane model with a horizontal fixed  
rail  at height $z=0$. A control trolley travels along the rail  holding a hoist rope of controllable length
$l$ from $x=0$ to $x=d$.  
We assume that the rope is rigid  for a  given length and also neglect its mass and damping, 
which are all standard approximations.   
Suppose that a point payload of mass $m$ is to be moved from $0,z_0$ to  $d,z_f$, 
with $z_0,z_f<0$, in a time $t_f$. 
(We shall in fact allow for some deviation from the ideal, equilibrium conditions.) 

For the  rectilinear motion of the suspension  point, transversal and longitudinal motions of the payload are uncoupled, so the dynamical problem is reduced to a 2-dimensional vertical plane with coordinates $\{x,z\}$.
Specifically we assume that the initial and final rope lengths are chosen as
\beq
l(0)=l_0=-z_0,\;\; l(t_f)=l_f=-z_f.
\label{lbc}
\eeq
The Lagrangian for the payload 
is, in terms of the swing angle $\theta$,
\beqa
L&=&\frac{m}{2}\left[\dot{x}^2+\dot{l}^2+l^2\dot{\theta}^2+2\dot{x}\dot{l}\sin \theta+2\dot{x} l\dot{\theta}\cos{\theta}\right]
\nonumber\\
&+&mgl\cos{\theta},
\label{Lagra}
\eeqa
where the first line is the kinetic energy ${\mathcal{T}}$, and the second line is minus the potential energy,
$-{\mathcal{V}}$. 
The corresponding dynamics, considering $l(t)$ and $x(t)$ as external control functions, is given by   
\beq
\label{dynam}
l\ddot{\theta}+2\dot{l}\dot{\theta}+g\sin\theta+\ddot{x}\cos\theta=0,
\eeq
where one or two dots denote first and second order time derivatives, and $g$ is the gravitational acceleration. 
Using as a new variable the horizontal deviation of the payload from the position of the trolley, $q=l\sin\theta$, 
and small oscillations, the 
kinematics of the load  are described by the linear equation 
\beq
\ddot{q}+\left(\frac{g}{l}-\frac{\ddot{l}}{l}\right)q=-\ddot{x},
\label{kinematics}
\eeq
which corresponds to a forced harmonic oscillator with squared, time-dependent angular frequency
\beq
\omega^2(t)=\frac{g}{l}-\frac{\ddot{l}}{l}. 
\label{del}
\eeq
We shall assume for a smooth operation that 
\beq
\dot{l}(t_b)=\ddot{l}(t_b)=0,
\label{lbc2}
\eeq
where we use the shorthand notation $t_b=0,t_f$ 
for the (initial or final) boundary times.  
In particular, the vanishing second derivative implies, see Eq. (\ref{del}),  that 
\beq
\omega^2(0)= \frac{g}{l_0}, \;\;
\omega^2(t_f)= \frac{g}{l_f}. 
\eeq
In a moving frame the kinematics in Eq. (\ref{kinematics}) may as well be derived from the Hamiltonian 
(The canonical transformations are given in Appendix B)
\beq
H=\frac{p^2}{2m}+\frac{m}{2}\omega^2(t)q^2+m\ddot{x} q,
\label{Hami}
\eeq
with $p=mdq/dt$,   
which has the invariant of motion \cite{LL} 
\beq
I=\frac{1}{2m}[b(p-m\dot{\alpha})-m\dot{b}(q-\alpha)]^2
+\frac{m}{2}\omega_0^2\left(\frac{q-\alpha}{b}\right)^2,
\label{invariant}
\eeq
provided the scaling factor $b(t)$ and $\alpha(t)$ satisfy the Ermakov and Newton equations 
\beqa
\ddot{b}+\omega^2(t)b&=&\frac{\omega_0^2}{b^3},
\label{Ermakov}\\
\ddot{\alpha}+\omega^2(t)\alpha&=&-\ddot{x},
\label{Newton}
\eeqa
where $\omega_0$ is in principle an arbitrary constant, but  it is customary and convenient to take 
$\omega_0=\omega(0)$. Note that Eqs. (\ref{kinematics}) and (\ref{Newton}) have
the same structure, corresponding to a forced oscillator. 
However, while Eq. (\ref{kinematics}) is general, and applies to arbitrary
boundary conditions for the trajectory, 
we shall choose $\alpha(t)$ functions that satisfy the boundary conditions 
\beq
\alpha(t_b)=\dot{\alpha}(t_b)=\ddot{\alpha}(t_b)=0.
\label{alphabc}
\eeq
%
These boundary conditions  also imply, see Eq. (\ref{kinematics}), that $\ddot{x}(t_b)=0$.  
Moreover, for the boundary conditions of the scaling function we choose
\beqa
b(0)&=&1, b(t_f)=\gamma=\left(\frac{\omega_0}{\omega_f}\right)^{1/2},
\nonumber\\
\dot{b}(t_b)&=&\ddot{b}(t_b)=0,
\label{rhobc}
\eeqa
where $\omega_f=\omega(t_f)$. 
$I(0)$ is equal to the initial energy of the payload $E_0=p^2(0)/(2m)+\frac{m}{2}\omega_0^2 q^2(0)$, 
where $q(0)$ and $p(0)$ are arbitrary initial conditions. 
$I(t_f)$, 
which must be equal to $E_0$ since $I$ is an invariant, 
takes the form $I(t_f)=\gamma^2 E_f$, where $E_f=p^2(t_f)/(2m)+\frac{m}{2}\omega_f^2 q^2(t_f)$ is the final energy, 
and $q(t_f)$ and $p(t_f)$ are found by solving Eq. (\ref{kinematics}) with the initial conditions $q(0)$, $p(0)$.
(These energies correspond to the Hamiltonian (\ref{Hami}), with zero potential at the equilibrium position 
$\theta=0$ or $q=0$. $H(t)$ is not in general equal to the shifted mechanical  
energy ${\mathcal{T}}+{\mathcal{V}}+mgl$, except at the boundary times.)  
In other words, for the processes defined by the auxiliary functions $\alpha(t)$ and $b(t)$ satisfying their stated
boundary conditions,
the final energy at time $t_f$ is, for any initial energy $E_0$,  the adiabatic energy $E_{ad}=E_0\omega_f/
\omega_0$, i.e., 
the one that could be  reached after  a slow evolution of the control parameters along an infinitely
slow process. This result is very relevant since, as shown in Appendix A, the minimal final average energy, 
averaged over an  initial microcanonical ensemble 
(a distribution of initial conditions proportional to Dirac's delta $\delta(E-E_0)$), 
is given by the adiabatic energy.     
  
To inverse engineer the control parameters  for a generic transport goal which involves simultaneous 
hoisting or lowering of the rope and  trolley transport we proceed as follows: 

{\it i})  $b(t)$
is interpolated between 0 and $t_f$ leaving two free parameters. 
A simple choice is the polynomial form 
\beqa
b(t)&=&\sum_{j=0,7}{a_j}S^j
\nonumber\\
&=&1+(-10+10\gamma-a_6-3 a_7)S^3
\nonumber\\
&+&(15-15\gamma+3 a_6+8 a_7)S^4
\nonumber\\
&+&3(-2+2\gamma-a_6-2a_7)S^5
\nonumber\\
&+&a_6S^6+a_7 S^7,
\eeqa
where the $a_j,\; j<6$, are  fixed by the boundary conditions (\ref{rhobc}), and $S\equiv t/t_f$. 

{\it ii}) The corresponding squared frequency $\omega^2(t; a_6,a_7)$ is deduced from the Ermakov equation (\ref{Ermakov}).

{\it iii}) The two free parameters in $b(t;a_6,a_7)$ are fixed by solving Eq. (\ref{del}) with initial conditions $l(0)=l_0,\, \dot{l}(0)=0$ ($\ddot{l}(0)=0$ is  automatically satisfied according to Eq. (\ref{del}))  so that the final conditions are $l(t_f)=l_f,\,\dot{l}(t_f)=0$ (again $\ddot{l}(t_f)=0$ is satisfied automatically). Multiple solutions are in principle possible because the system is non-linear. We use a root finding subroutine (FindRoot by MATHEMATICA) starting with the seed 
$a_6=a_7=0$. This specifies the form of $l(t)$ and $\omega(t)$. 

{\it iv}) A functional form for $\alpha(t)$ is used satisfying the boundary conditions (\ref{alphabc}) with two free 
parameters. Again, a polynomial is a simple, smooth choice,
\beqa
\alpha(t)&=&\sum_{j=0,7} b_j S^j
\nonumber\\
&=&-(b_6+3b_7)S^3+(3b_6+8b_7)S^4
\nonumber\\
&-&3(b_6+2b_7)S^5+b_6S^6+b_7 S^7.
\label{alphat}
\eeqa
The trajectory of the trolley is
deduced from Eq. (\ref{Newton}) as 
\beq
x(t) = - \int^t_0 dt' \int^{t'}_0 dt'' [\ddot{\alpha}(t'') + \omega ^2 (t) \alpha ],
\label{trajectory}
\eeq
which satisfies $\ddot{x}(t_b)=\dot{x}(0)=x(0)=0$.
The two free parameters in $\alpha(t;b_6,b_7)$ are set by demanding
\beq
x(t_f)=d, \dot{x}(t_f)=0.
\label{bcxtf}
\eeq
Compare the above sequence to a simpler inverse approach in which $l(t)$  is designed to satisfy Eqs. (\ref{lbc}) 
and (\ref{lbc2}) 
so as to get $\omega(t)$ from Eq. (\ref{del}), and then 
$\alpha(t)$ is designed to satisfy the conditions in Eq. (\ref{alphabc}).   In this simpler approach  $b(t)$ is not engineered, 
which means that in general it will not satisfy  the boundary conditions in Eq. (\ref{rhobc}), and, as a consequence, a   
vanishing residual excitation (with respect to the adiabatic energy) is only guaranteed 
for a $q(t)$ that satisfies the initial boundary conditions in Eq. (\ref{alphabc}). In other words, the extra effort to 
design $b(t)$ leads by construction, see Eq. (\ref{invariant}), 
to the adiabatic energy at final time for any initial boundary conditions  $q(0),\,\dot{q}(0)$.

%
%
%
%
%
%
\section{Numerical examples\label{ne}}
%
\begin{figure}[t]
\begin{center}
\includegraphics[width=8.cm]{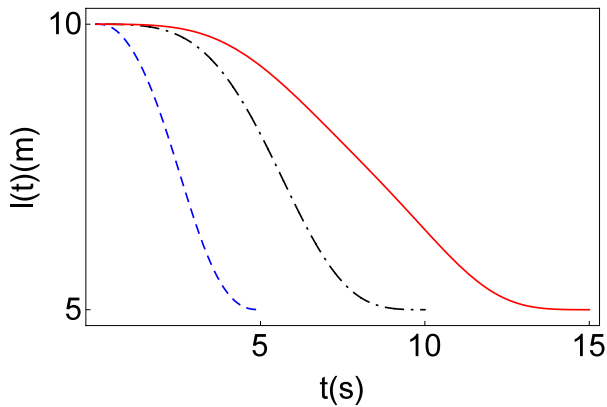}
\caption{(Color online) Length of the rope with respect to time for a hoisting from $l_0=10$ m to $l_f= 5$ m. $t_f=15$ s (red solid line); $t_f=10$ s (black dash-dotted line); $t_f=5$ s (blue dashed line).}
\label{lfig}
\label{angles}
\end{center}
\end{figure}
\begin{figure}[t]
\begin{center}
\includegraphics[width=8.cm]{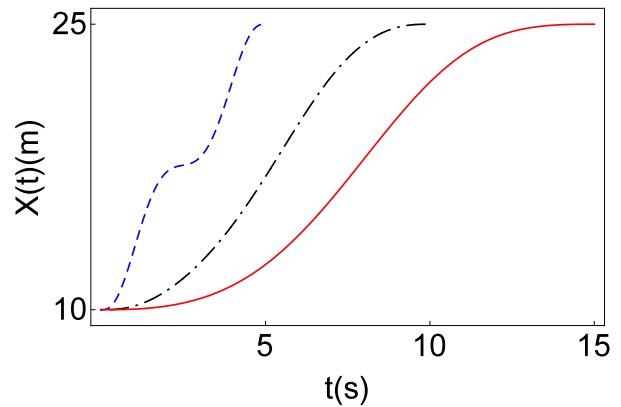}
\caption{(Color online) Position of the trolley $x(t)$ with respect to time for a dual operation with $d=15$, $l_0=10$ m, and  $l_f= 5 m$. $t_f=15$ s (red solid line); $t_f=10$ s (black dash-dotted line); $t_f=5$ s (blue dashed line).}
\label{xfig}
\label{angles}
\end{center}
\end{figure}
%
%
%
%
\begin{figure}[t]
\begin{center}
\includegraphics[width=8.cm]{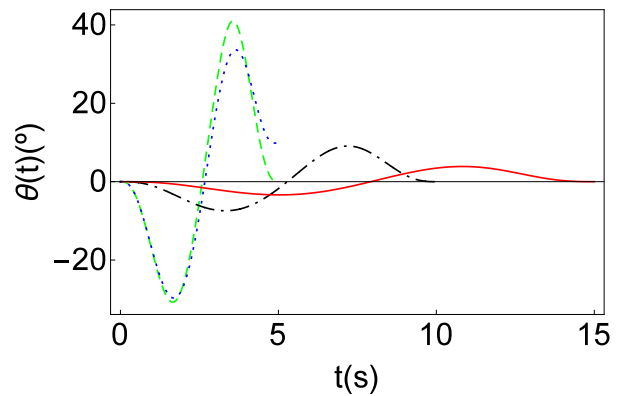}
\caption{(Color online) Angle of the payload with respect to time for a dual operation with a traslation of $15$ m
and hoisting from $l_0=10$ m to $l_f= 5$ m).  Exact dynamics: $t_f=15$ s (red solid line); 
$t_f=10$ s (black dash-dotted line); $t_f=5$ s (blue dotted line). Approximate dynamics (harmonic approximation):   $t_f=5$ s (green dashed line); for the two other times the exact and approximate curves are  undistinguishable
in the scale of the figure.   Payload initially at equilibrium.}
\label{fig:ejemplo}
\label{angles}
\end{center}
\end{figure}
\begin{figure}[t]
\begin{center}
\includegraphics[width=8.cm]{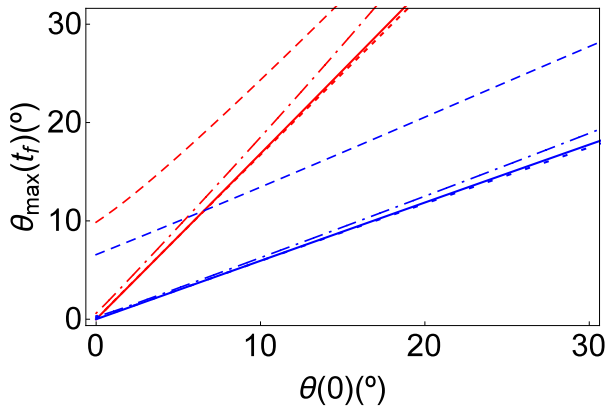}
\caption{(Color online) Maximal angle of the payload in the final configuration versus the initial angle for the payload
initially at rest $\dot{\theta}(0)=0$.
For the same line type, the (red) curve with a larger slope  corresponds to transport with hoisting ($l_0=10$ m, $l_f=5$ m, $d=15$ m) 
and the other (blue) curve  to transport with  lowering ($l_0=5$ m, $l_f=10$ m, $d=15$ m). 
The continuous lines  are the (exact) adiabatic lines. $t_f=10$ s (dotted lines); $t_f=7$ s (dash-dotted lines); 
$t_f=5$ s  (dashed lines).}
\label{errors}
\end{center}
\end{figure}

%
In this section we show some examples of crane operation and payload behavior following the inverse engineering protocol
described in steps {\it i} to {\it iv}.  
Figures \ref{lfig} and \ref{xfig} show the control functions  $l(t)$ and $x(t)$ found for different final times, with parameters $d=15$ m, 
$l_0=10$ m, and $l_f=$ 5 m.  
Figure \ref{angles} shows the swing angle of the payload with respect to time for these protocols, with the payload initially at equilibrium. It is calculated exactly (with the full Lagrangian) for different final times, and also with the approximate equation for small oscillations (\ref{kinematics}), but the difference is only noticeable for the smallest time $t_f=5$ s.
Clearly short times imply larger transient angles so that larger errors may be expected as the small oscillations regime is abandoned. 
To quantify the error we plot in Fig. \ref{errors} the maximal angle of the payload in the final configuration, 
$\theta_{max}(t_f)$, i.e. the maximal angle for the
pendulations with $l=l_f$ once the trolley has reached the final point $d$ and remains at rest, 
versus the initial angle (with $\dot{\theta}(0)=0$).
The maximal angle is of course related to the exact final energy, given  
(since  $\dot{l}(t_f)=\dot{x}(t_f)=0$)   
%
%
by  $E_f=mgl(1-\cos[\theta_{max}(t_f)])$.   
In Fig. \ref{errors}, we compare this final maximal angle to the maximal angle that would be found adiabatically after an infinitely slow process. 
The exact adiabatic angle can be calculated  as explained in Appendix C. 
The figure demonstrates that hoisting the payload (cable shortening) leads to  
larger maximal angles than when lowering it. 
      
%
%
\subsection{Sequential versus dual}
%
%
\begin{table}
\centering
\scalebox{0.78}{
\begin{tabular}{|c|c|c|c|c|c|c|c|}
\hline
\multirow{2}{*}{d(m)}& \multirow{2}{*}{$\Delta$l(m)}
&\multicolumn{5}{c|}{Sequential time (s)} 
& \multirow{2}{*}{Dual time (s)}
\\
   \cline{3-7}
   &  & \multicolumn{2}{c|}{Transport}  &   {Lowering} &\multicolumn{2}{c|}{Total}&
\\
\hline
{15}& {5}
& \multicolumn{2}{c|}{9.1$^{(1)}$}    &  {1.29$^{(2)}$} &  \multicolumn{2}{c|}{10.39} & 9.55$^{(1)}$
\\
\hline
{15}& {30}
&\multicolumn{2}{c|}{9.1$^{(1)}$} &  {1.16$^{(2)}$} &   \multicolumn{2}{c|}{10.26}&10.7$^{(1)}$
\\
\hline
{50}& {30}
&\multicolumn{2}{c|}{16.6$^{(1)}$} & {1.16$^{(2)}$} & \multicolumn{2}{c|}{17.76}&21$^{(1)}$
\\
\hline
{5}& {50} & 5.25$^{(1)}$$_{(l_0)}$ & 7.76$^{(3)}$$_{(l_f)}$ & 1.1$^{(2)}$&6.35$_{(l_0)}$ & 8.86$_{(l_f)}$ &{7$^{(1)}$ } 
\\
\hline 
\end{tabular}}
\caption{Minimal times for sequential and dual protocols with different parameters.  $\Delta l=l_f-l_0$ with $l_0=5$m for all cases. Times for the sequential process are divided into minimal times for pure transport (in red) and pure lowering (in blue) processes. Superscripts indicate the constraint that sets the minimal time, standing for $(1):$ $-10 \leq \theta (t) \leq 10$, $(2):$ $0 \leq l(t) \leq l_f$, and $(3):$ $0 \leq x(t) \leq d$. The transport in the last row has two minimal times because pure transport with $l_0$  and with $l_f$  are bounded by different constraints. In all other cases the constraint is the same for both lengths, so they share the same minimal time, see main text.\label{table1}}
\end{table}
A global process that includes transport and hoisting/lowering can be done performing the two operations simultaneously, as described above (dual operation), but also sequentially, performing one operation at a time \cite{xia}. Which one of the two possibilities is faster? 
For the analogous problem consisting on atom transport or launching combined with trap expansions, we have found that the answer depends critically on the constraints and parameters imposed  \cite{ander}.
In the current setting, we shall also demonstrate with examples that, by imposing some constraints, either the dual or the sequential approach
could be faster depending on the parameter values chosen. 
A dual protocol is not just the simultaneous superposition of  the two sequential operations: While $l(t)$ is indeed the same in the dual and 
hoisting/lowering  part of the sequential process   (the first three steps, {\it i}, {\it ii}, {\it iii}, 
would be identical), $x(t)$ is different in the dual process and in the 
pure transport segment of the sequential process, where $l$ is constant.    
This is because, in  Eq. (\ref{Newton}), the time dependence of the angular frequency affects the design of $\alpha(t)$ and $x(t)$.

Consider  processes where the trolley moves from $0$ to $d$ with lowering from $l_0$ to $l_f>l_0$. 
We assume initial conditions $q(0)=\dot{q}(0)=\ddot{q}(0)=0$ (which implies $q(t)=\alpha(t)$, see Eqs. (\ref{kinematics}) and (\ref{Newton})),
and protocols for the control functions using polynomial ansatzes as described above. 
We impose three different constraints: $-10 \leq \theta(t) \leq 10$ may be imposed for a safe operation and guarantees the validity of the harmonic model;   $0 \leq x(t) \leq d$, and $0 \leq l\leq l_f$ are assumed geometrical constraints on the trolley trajectory and cable length.
Of course other geometries may as well be considered. For example, the presence of obstacles may imply the necessity of a sequential approach. 
%

%


Table \ref{table1} shows the minimal times found for different cases. In all,  $l_0=5$ m and $l_f>l_0$, and the protocols are 
found according to the steps described above. 
The fastest protocol may be dual or sequential depending on the parameters chosen. 
Except in the last line, the minimal time for the sequential approach is the same regardless of the order in which the two 
operations (pure transport and pure lowering) are performed.  In other words, the minimal time for transport is the same for $l_0$ or $l_f$. 
Notice that for a constant $\omega$, the free parameters in $\alpha(t;b_6,b_7)$,
fixed so that boundary conditions in Eq. (\ref{bcxtf}) are satisfied, are explicitly given by
\beq
b_6= \frac {17640 d}{t_f^2 \omega^2},\, \,\,\,b_7= - \frac{5040 d}{t_f^2 \omega^2},
\eeq
which yields the scaling $\omega_0^2 \alpha_0(t) = \omega_f^2 \alpha_f(t)$, where  $\alpha_0(t)$ and $\alpha_f(t)$ correspond to transport with $l_0$ and $l_f$ respectively. 
Using Eq. (\ref{del}) (with $\ddot{l}=0$) it is found that 
\beq
\frac{\alpha_0(t)}{l_0}=\frac{\alpha_f(t)}{l_f}
\label{compa2},
\eeq
which means that the triangle formed by the cable of length $l_0$, the horizontal displacement $\alpha_0(t)$,  and the vertical, is similar to the triangle formed by $l_f$,  $\alpha_f(t)$ and the vertical. Thus, the evolution of the angle $\theta(t)$, is independent of the rope length in these protocols.

This symmetry is broken in the last line of Table \ref{table1} because the transport function with $l_f$ similar to the transport process 
that gives the minimal time with $l_0$, violates the constraint imposed on $x(t)$. We have then to increase the process time until the constraint is
satisfied, so that Eq. (\ref{compa2}) does not apply.       


%
%

%
%
%
%
%
%
%
\section{Discussion}
\label{sec4}
%
%
\begin{figure}[t]
\begin{center}
\includegraphics[width=8.cm]{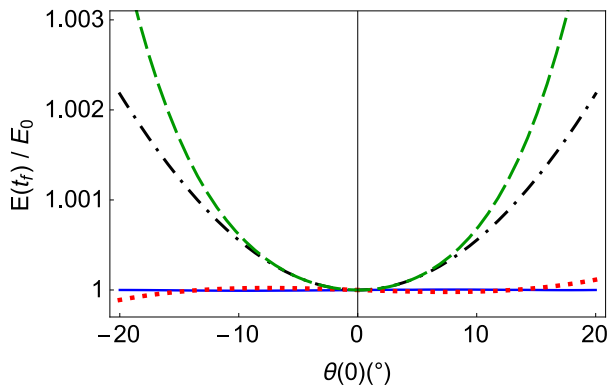}
\caption{(Color online) Final to initial energy ratio for pure transport (no hoisting involved) versus initial angle  
for load initially at rest. Here the initial energy is equal to the final adiabatic energy as the cable length does not change. Protocols:   
{\it i}) Three-step protocol  in \cite{sun} in the harmonic approximation (black dot-dashed line), 
and for the exact dynamics (green dashed line);  {\it ii}) Invariant-based protocol (red dotted line) with  
polynomial ansatz (\ref{alphat}) for $\alpha$ (blue solid line in the harmonic approximation; red dotted line for the exact dynamics);  
{\it iii}) Invariant-based protocol with three more parameters in the polynomial adjusted  to flatten the excitation curve
as in \cite{energy} (blue solid line, indistinguishable from 1 in the scale of the figure throughout the whole angle interval).
Parameters taken from \cite{sun}: $l=1.2$ m and $x_f=4$ m. Constraints to set the protocol {\it{i}}): $a_{ub}=0.5$ m/s$^2$, $v_{ub}=1$ m/s, $\theta_{ub}=5^\circ$. This implies a total time $t_f=6.45$ s with  
$a_{max}=0.4276$ m/s$^2$, $T=2.1987$ s, and $t_c=2.0567$ s. All protocols are adjusted for the same final total time.}
\label{sunfig}
\end{center}
\end{figure}
%
%
Analogies among disparate systems provide opportunities for a fruitful interchange of techniques and results. 
Here we have worked out an invariant-based inverse engineering approach to control crane operations, which had been 
applied so far to control the transport of atoms and ions.  
We provide protocols that guarantee final adiabatic energies, which are shown to  be minimal when averaging over a microcanonical ensemble of initial conditions. Natural applications of these protocols would be in robotic cranes with good control 
of the driving parameters and uncertainties in the initial conditions. 
   
  
Several results in the microscopic domain may be translated to crane control. Here are some examples of possible connections
for future work: 
 
- In \cite{PRA2014} the excess final energy in transport processes was related to a Fourier transform of the acceleration of the trap. 
By a clever choice of acceleration function it is possible to set a robustness window for the trap frequency.
In the context of cranes, this would provide robustness with respect to rope length in transport operations.  

- A more realistic treatment of the crane implies a double pendulum \cite{swing}. 
Setting the shortcuts in that case will require considering two harmonic oscillators, by means of a dynamical normal mode analysis 
similar to the one done for two ions \cite{Lizuain}. 

- Invariant-based inverse engineering combines well with Optimal Control Theory \cite{invoct1,invoct2,invoct3}. The main point is that the 
shortcut design guarantees a final excitation-free state, but leaves room to choose the control parameters so as to optimize some relevant variable, 
typically for a bounded domain of values for the control parameters. Results found for microscopic systems to minimize times or transient energies, can be applied to crane systems.  
  
- Noise and perturbations, including anharmonicities may be treated perturbatively to minimize their effect, as in \cite{NJP2012,anhar}.

An alternative practical scheme for dealing with anharmonicity (i.e., non-linear effects beyond the small-oscillations regime) was put forward in \cite{energy}: Instead of using a minimum set of parameters to design of the auxiliary functions 
($\alpha$ and $b$), ansatzes with additional parameters enable us to minimize the final excitation for a broad domain of
initial angles. In Fig. \ref{sunfig} we depict ratios of final to initial energies for a pure transport process. 
One of the protocols used was devised in \cite{sun} in the harmonic regime to let the load at rest  
at final time when starting at 
$\theta(0)=0$, $\dot{\theta}(0)=0$ by three steps with constant (stepwise) acceleration:  $a_{max}$ during an oscilation period $T$, coasting (constant velocity) for $t_c$, and $-a_{max}$ for a time $T$. $a_{max}$ and $t_c$ are chosen to satisfy imposed  bounds on the swing angle, the maximal velocity $a_{ub}$, and maximum velocity $v_{ub}$ for the trolley. 
The values, taken from an example in \cite{sun},  are given in the caption. 
The ratio grows from one as the initial angle increases, even in the harmonic approximation   
(black dot-dashed line). The exact results give even higher values (green dashed line). 
By contrast, the invariant-based protocol using $b(t)=1$, and the polynomial ansatz for $\alpha$, Eq. (\ref{alphat}),
with parameter values  to satisfy 
Eqs. (\ref{alphabc}) and (\ref{bcxtf}), gives no  excitation at all in the harmonic approximation 
(this is so by construction, see the blue solid line), and little excitation when using the exact dynamics
(red dotted line). The result can be made even better by increasing the order of the polynomial for $\alpha$ and 
using the free parameters to minimize the total excitation at a grid of selected initial angles, see \cite{energy}.
Choosing three more parameters and the angles $\theta(0)=-20, -15, 15, 20$, the energy ratio becomes indistinguishable
from 1.          
   
We end with an important example of a positive influence in the opposite direction, from the macroscopic to the microscopic realm.  
The analysis of energy management and expenditure has been more realistic in engineering publications, see e.g. \cite{xia}, than in the 
work on microscopic systems. A useful energy cost analysis of a shortcut must include the control system (trolley) in addition to the primary system (payload). 
In \cite{energy}, the dynamics of the crane is analyzed in terms of 
coupled dynamical equations for the trolley and the payload \cite{xia} for a transport operation (with constant $l$), 
\beqa
0&=&l\ddot\theta+\ddot x\cos\theta+g\sin\theta,
\label{dyn1}
\\
\mathcal{F}_a+\mathcal{F}_r&=&M\ddot x+m(\ddot x+l\ddot\theta\cos\theta-l\dot\theta^2\sin\theta), 
\label{dyn2}
\eeqa
where $M$ is the mass of the trolley, $\mathcal{F}_a(t)$ an actuating force (due to an engine or braking),
and $\mathcal{F}_r$ a friction force, simply modeled as $\mathcal{F}_r=-\Gamma\dot{x}$, $\Gamma\ge 0$.
In contrast to Eq. (\ref{dynam}), the trolley position is now regarded as a dynamical variable. To achieve a  
given form $x(t)$, as specified by the inverse engineering approach, $\mathcal{F}_a(t)$ must be modulated accordingly, 
so it will depend on the specified $x(t)$, on trolley characteristics (mass $M$ and friction coefficient $\Gamma$), and payload dynamics 
and characteristics.    
The power due to the actuating force,  $\mathcal{P}=\mathcal{F}_a\dot x$, may be easily translated into fuel, or electric power
consumption, 
and is in general quite different from the power computed as the time derivative of the mechanical energy of the payload alone.  
It takes for small oscillations the simple form 
\beqa
\label{Preal}
\mathcal{P}&=&\mathcal{F}_a\dot x=\left(M\ddot x-mq \omega^2+\Gamma \dot{x}\right)\dot x,
\eeqa
whose peaks and time integrals are studied in \cite{energy}, with different treatments for different braking mechanisms.
The energy cost of shuttling neutral atoms in  optical traps moved by 
motorized motions of mirrors or lenses \cite{Couvert, Oliver} may in fact be analyzed in the same manner,
as the shortcut design is based on exactly the same equations.    
\section*{Acknowledgments}
We acknowledge discussions with S. Mart\'\i nez-Garaot and M. Palmero.  
This work was supported by 
the Basque Country Government (Grant No.
IT986-16); 
MINECO/FEDER,UE (Grants No.
FIS2015-67161-P and FIS2015-70856-P);  QUITEMAD+CM S2013-ICE2801; and by Programme Investissements d'Avenir under the program ANR-11-IDEX-0002-02, reference ANR-10-LABX-0037-NEXT. 
%
%
%
\appendix
\section{Minimal work\label{aa}}
Here we follow closely Ref. Ê\cite{adiabinv}, the main difference is that we allow for a harmonic Hamiltonian with a 
homogeneous, time-dependent force $F(t)$, 
\beq
H=p^2/2m+\frac{m}{2}\omega^2(t)q^2-F(t)q,
\eeq
which we assume to satisfy $F(0)=F(t_f)=0$. 
Consider the area preserving phase-flow map between initial and final values of position and momentum,  
\beqa
\Phi &:& \left(\begin{array}{c}q_0\\p_0\end{array}\right)\to 
\left(\begin{array}{c}q_f\\p_f\end{array}\right),
\\
\Phi&=&\left(\begin{array}{cc}a&b\\c&d\end{array}\right),
\eeqa
where $\det(\Phi)=1$. In terms of the matrix elements of $\Phi$, the final energy for initial conditions $q_0$, $p_0$ with energy $E_0$ is   
\beq
E_f=E_0(\alpha\cos^2\phi+\beta\sin^2\phi+\eta\sin 2\phi),
\eeq
where 
\beqa
\alpha&=&\frac{c^2}{m^2\omega_0^2}+a^2\frac{\omega_f^2}{\omega_0^2},\;
\beta=d^2+\omega_fm^2b^2,
\\
\eta&=&\frac{cd}{m\omega_0}+abm\frac{\omega_f^2}{\omega_0},
\eeqa
and the angle $\phi$ is introduced as 
\beq
q_0=\sqrt{\frac{2E_0}{m\omega_0^2}}\cos\phi,\;\;p_0=\sqrt{2mE_0}\sin\phi.
\eeq
For a uniform (microcanonical) distribution of initial angles, $P(\phi)=1/(2\pi)$, 
\beq
\overline{E}_f=\frac{1}{2\pi}\int E_f d\phi=\frac{E_0}{2} (\alpha+\beta).
\eeq
This gives for the variance 
\beq
\mu^2\equiv\overline{(E_f-\overline{E}_f)^2}=\frac{E_0^2}{2}\left[\left(\frac{\overline{E}_f}{E_0}\right)^2-\frac{\omega_f^2}{\omega_0^2}\right], 
\eeq
so $\overline{E}_f\ge E_0\frac{\omega_f}{\omega_0}$, i.e., the final averaged energy is greater or equal than the adiabatic energy at $t_f$, $E_{ad}=E_0\omega_f/\omega_0$. 
\section{Detailed derivation of the Hamiltonian}  
To derive the Hamiltonian (\ref{Hami}), let us first rewrite it in a slightly more detailed notation as
\begin{equation}
H_q=\frac{p_q^2}{2m}+\frac{m}{2}\left(\frac{g}{l}-\frac{\ddot l}{l}\right)q^2+m q \ddot x.
\label{rewr}
\end{equation}
{\it Lagrangian and Hamiltonian in $\theta$.}
The starting point is the Lagrangian of the system in the $\theta$ variable, Eq. (\ref{Lagra}). 
Let us define the conjugate momentum $p_\theta$ as
\begin{equation}
p_{\theta}=\frac{\partial L}{\partial \dot\theta}
 =ml^2\dot \theta+ml\dot x\cos\theta.
 \label{ptheta}
\end{equation}
%
We now define the Hamiltonian as the Legendre transform of the Lagrangian writing
everything in terms of $\theta$ and $p_\theta$,
\begin{eqnarray}
  H_\theta&=&\dot\theta p_{\theta}-L
\nonumber\\
&=&\frac{p_\theta^2}{2 m l^2}
-m g l \cos \theta
-\frac{ \dot x\cos \theta }{l}p_\theta
\nonumber\\
&-&\frac{1}{2}m \left(\dot x^2\sin^2\theta+2\dot l\dot x \sin\theta\right),
\label{Htheta}
 \end{eqnarray}
where we have omitted terms that do not depend on $\theta$ or $p_\theta$
so that they do not affect  the dynamics. 

{\it Horizontal deviation Q: change of coordinate.}
   We look for a canonical transformation to  $\{Q,p_Q\}$ variables such that the new coordinate is the horizontal deviation from the suspension point
   $Q=l\sin \theta$. This is achieved with the (time-dependent) generating function $F_2=p_Q l \sin\theta$. 
$F_2$ generates the desired change of coordinate since
\begin{eqnarray}
  Q&=&\frac{\partial F_2}{\partial p_Q}=l\sin \theta,
  \\
  p_\theta&=&\frac{\partial F_2}{\partial \theta}=p_Q l \cos\theta.
 \label{ptheta} 
 \end{eqnarray}
Including the inertial effects given by $\partial_t F_2=p_Q\dot l \sin\theta=p_Q Q\dot l/l$, the Hamiltonian in the new variables  is
 \begin{eqnarray}
  H_Q&=&\frac{p_Q^2}{2m}\left(1-\frac{Q^2}{l^2}\right)-mgl \sqrt{1-\frac{Q^2}{l^2}}
  -\frac{m  \dot l \dot x}{l}Q
 \nonumber\\
 &-&\frac{m  \dot x^2}{2 l^2}Q^2+\frac{\dot x}{l^2}p_Q Q^2
  +p_Q\left(\frac{\dot l}{l} Q-\dot x\right).
 \end{eqnarray}
%

{\it Momentum shift.}
The last term introduces a quadratic coordinate-momentum coupling  and  a linear-in-momentum term.
To get rid of this term we make a momentum shift by a second canonical transformation using the  generating function
 \begin{equation}
F'_2=Q p_q+m\dot x Q-\frac{m\dot l}{2l}Q^2.  
 \end{equation}
The transformation equations to the new canonical variables $\{q,p_q\}$ are 
\begin{eqnarray}
  q&=&\frac{\partial F'_2}{\partial p_q}=Q,\\
  p_Q&=&\frac{\partial F'_2}{\partial Q}=p_q+m\dot x-\frac{m\dot l}{l}Q.
  \label{pQ}
 \end{eqnarray}
 Including the inertial effects
 \begin{equation}
  \partial_t F'_2=-\frac{m Q^2 \ddot l}{2 l}+\frac{m Q^2 \dot l^2}{2 l^2}+m  \ddot x Q,
 \end{equation}
 the transformed Hamiltonian in $q$ and $p_q$ variables is given by
\beqa
  H_q&=&\frac{p_q^2}{2m}\left(1-\frac{q^2}{l^2}\right)-mgl \sqrt{1-\frac{q^2}{l^2}}
  +m \ddot x q
 \nonumber\\ 
    &-&\frac{m \ddot l}{2 l}q^2
  -\frac{m  \dot l^2}{2 l^4}q^4+\frac{ \dot l}{l^3}p_q q^3,
 \eeqa
where, again, terms that do not affect the dynamics have been omitted.

{\it Small oscillations.}
 In the $q\ll l$ limit, keeping only quadratic terms in coordinate and momentum and up to terms that 
 do not affect the dynamics, the above Hamiltonian may be approximated by Eq. (\ref{rewr}).
%
The next order is  quartic (there is no  cubic term),
\begin{eqnarray}
 V&=&\left(\frac{g m }{8 l^3}-\frac{m \dot l^2}{2 l^4}\right)q^4-\frac{p_q^2 q^2}{2 m l^2}+\frac{\dot l}{l^3}p_q q^3.
\end{eqnarray}
Using the relations between momenta $p_q$, $p_Q$ and $p_\theta$, see Eqs. (\ref{ptheta}, \ref{pQ}),  $p_q$ can be written  in terms of 
the original coordinates $\theta$ and $\dot \theta$,
\begin{equation}
 p_q=m \left(l \dot \theta \sec\theta + \dot l \sin\theta\right)\approx m\left(l \dot\theta +\dot l\theta\right)=m\frac{d(l\theta)}{dt}.
\end{equation}
In other words,  in the small oscillation regime where $q=l \sin\theta\approx l\theta$, 
the momentum $p_q$ tends to $p_q\approx m\dot q$.
\section{Exact adiabatic maximal angle\label{ab}}
We shall determine the exact adiabatic maximal final angle of the payload. ``Exact'' here means that we do not use the harmonic approximation to calculate it. It  would be the  maximal angle for the final cable length  and some given initial length and energy, after a very slow process.  The maximal angle depends on the energy, so we first need an exact expression of the energy at the boundary times for an arbitrary process where the boundary conditions $\dot{l}=\dot{x}=\ddot{l}=\ddot{x}=0$ are imposed. Using the exact Lagrangian (\ref{Lagra}),  and $\theta$ as the 
generalized coordinate, so that $p_\theta=\partial L/\partial \dot{\theta}$, we find by a Legendre transformation  
\beq
\label{ade}
\widetilde{E}=\frac{p_\theta^2}{2ml^2}-mgl\cos\theta,
\eeq
where the tilde in $\widetilde{E}$ indicates that the zero of the potential  energy corresponds here,
unlike the definition in the main text based on Eq. (\ref{Hami}), to 
$\theta=\pi/2$, and $l=l_0,l_f$.
The maximum angle $\theta_{max}$ for the boundary configurations, not to be confused with the angles
at boundary times $\theta(0)$ or $\theta(t_f)$, 
is given by 
\beq\label{cosE}
\cos \theta_{max}=-\frac{\widetilde{E}}{mgl}.
\eeq
The adiabatic invariant \cite{Gold},   
which is the phase-space area
along an oscillation cycle,  with $l$ fixed, is given 
at the boundary times by  
\beqa
&&A(l,\theta_{max})=4\!\!\int_0^{\theta_{max}}\! \!p_\theta d_\theta
\nonumber\\
&&=
4\sqrt{2m}\, l\! \int_0^{\theta_{max}}\!\! \sqrt{\widetilde{E}+mgl\cos \theta}\, d\theta
\nonumber\\
&&= 8\sqrt{2} m l^{3/2} \sqrt{1-\cos \theta_{max}} {\bf{E}}\!\left(\!\frac{\theta_{max}}{2}\Bigg|{\frac{2}{1-\cos \theta_{max}}}\!\right)\!,
\nonumber\\
\eeqa
where ${\bf{E}}$ is the incomplete elliptic integral of second kind \cite{elliptic,Gradstein}.
The exact final adiabatic maximum angle
$\theta_{max}(t_f)$ is found by imposing the equality of areas at the boundary times,   
%
\beq
\label{areas}
A[l_0,\theta_{max}(0)]=A[l_f,\theta_{max}(t_f)], 
\eeq
where the initial maximum angle $\theta_{max}(0)$ is
found from the initial energy via Eq. (\ref{cosE}). The final adiabatic energy can be also obtained from 
Eq. (\ref{cosE}).  
As a consistency check, using the following properties of the elliptic integral \cite{elliptic,Abra}
\beqa
{\bf{E}}(u|v)&=&v^{1/2}{\bf{E}}(uv^{1/2}|v^{-1})-(v-1)u,
\\
{\bf{E}}(z|x)&\sim& z-\frac{2z-\sin(2z)}{8}x\;\;  {\rm as}\, x\sim 0,  
\eeqa
and the alternative convention for the zero of the energy, $E=\widetilde{E}+mgl$, 
Eq. (\ref{areas}) leads to the expected result $E_f=E_0\omega_f/\omega_0$ in the limit of small oscillations.  
For an arbitrary  process, the actual final maximal angle and corresponding energy will differ from the adiabatic ones, 
as Fig. \ref{errors} demonstrates.


\end{document}